\PassOptionsToPackage{table}{xcolor}
\documentclass[10pt,conference]{IEEEtran}
\IEEEoverridecommandlockouts

\usepackage{cite}
\usepackage{amsmath,amssymb,amsfonts}
\usepackage{algorithmic}
\usepackage{graphicx}
\usepackage{textcomp}
\usepackage{xcolor}
\usepackage{lscape}
\usepackage{tcolorbox}
\usepackage{graphics}
\usepackage{booktabs}
\usepackage{amsmath,amssymb,amsfonts}
\usepackage{algorithmic}
\usepackage{graphicx}
\usepackage{textcomp}
\usepackage{xcolor}
\usepackage{wasysym}
\usepackage{multirow}
\usepackage{makecell}
\usepackage{subcaption}
\usepackage{enumitem}
\usepackage{listings}
\usepackage{soul,color}
\usepackage{subcaption}
\usepackage{hyperref}
\usepackage{soul}
\usepackage[framemethod=TikZ]{mdframed}
\usepackage{wrapfig}
\usepackage{flushend}

\def\BibTeX{{\rm B\kern-.05em{\sc i\kern-.025em b}\kern-.08em
    T\kern-.1667em\lower.7ex\hbox{E}\kern-.125emX}}

\hypersetup{
    linkcolor=blue,
}

\begin{document}

\title{\textsc{SeePrivacy}: Automated Contextual Privacy Policy Generation for Mobile Applications}

\author{
    \IEEEauthorblockN{
    Shidong Pan\IEEEauthorrefmark{1}\IEEEauthorrefmark{2}\IEEEauthorrefmark{4}, 
    Zhen Tao\IEEEauthorrefmark{1}, 
    Thong Hoang\IEEEauthorrefmark{2}, 
    Dawen Zhang\IEEEauthorrefmark{1}\IEEEauthorrefmark{2}, Zhenchang Xing\IEEEauthorrefmark{1}\IEEEauthorrefmark{2}, Xiwei Xu\IEEEauthorrefmark{2}, Mark Staples\IEEEauthorrefmark{2}}
    \IEEEauthorblockN{David Lo\IEEEauthorrefmark{3}}
    \IEEEauthorblockA{\IEEEauthorrefmark{1}Australian National University, Canberra, Australia
    \\\{Shidong.Pan, Zhen.Tao\}@anu.edu.au}
    \IEEEauthorblockA{\IEEEauthorrefmark{2}CSIRO's Data61, Sydney, Australia
    \\\{James.Hoang, Dawen.Zhang, Zhenchang.Xing, Xiwei.Xu, Mark.Staples\}@data61.csiro.au}
    \IEEEauthorblockA{\IEEEauthorrefmark{3}Singapore Management University, Singapore
    \\\{DavidLo\}@smu.edu.sg}
}




\maketitle

\begingroup
\renewcommand\thefootnote{\IEEEauthorrefmark{4}}
\footnotetext{The work was mostly completed during the academic visit at Singapore Management University.}
\endgroup

\pagestyle{plain}

\begin{abstract}
%
Privacy policies have become the most critical approach to safeguarding individuals’ privacy and digital security. To enhance their presentation and readability, researchers propose the concept of contextual privacy policies (CPPs), aiming to fragment policies into shorter snippets and display them only in corresponding contexts. In this paper, we propose a novel multi-modal framework, namely \href{https://cpp4app.github.io/}{\textsc{SeePrivacy}} 
, designed to automatically generate contextual privacy policies for mobile apps. Our method synergistically combines mobile GUI understanding and privacy policy document analysis, yielding an impressive overall 83.6\% coverage rate for privacy-related context detection and an accuracy of 0.92 in extracting corresponding policy segments. Remarkably, 96\% of the retrieved policy segments can be correctly matched with their contexts. The user study shows SeePrivacy demonstrates excellent functionality and usability (4.5/5). Specifically, participants exhibit a greater willingness to read CPPs (4.1/5) compared to original privacy policies (2/5). Our solution effectively assists users in comprehending privacy notices, and this research establishes a solid foundation for further advancements and exploration.
\end{abstract}


\maketitle
\section{Introduction}
\label{sec_intro}

Mobile apps have become an omnipresent part of our digital lives, with billions of downloads worldwide. These apps offer convenience, entertainment, and various services; however, they also collect and exchange vast amounts of personal information, raising increasing concerns about data privacy. For example, the British Information Commissioner's Office accused Tiktok of violating data protection for children.\footnote{\url{https://www.darkreading.com/vulnerabilities-threats/tiktok-other-mobile-apps-violate-privacy-regulations}} In 2021, Apple blocked 1.6 million mobile applications for misleading, spamming, and defrauding users.\footnote{\url{https://www.bleepingcomputer.com/news/security/apple-blocked-16-millions-apps-from-defrauding-users-in-2021/}}

%
\begin{figure} [t!]
\begin{subfigure}{.324\linewidth}
  \centering
  \includegraphics[width=.98\linewidth]{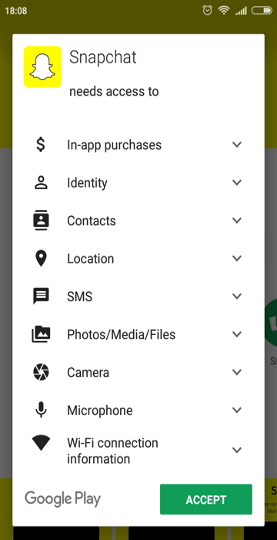}
  \caption{Install-time}
  \label{fig:sfig1}
\end{subfigure}%
\thinspace
\begin{subfigure}{.315\linewidth}
  \centering
  \includegraphics[width=.98\linewidth]{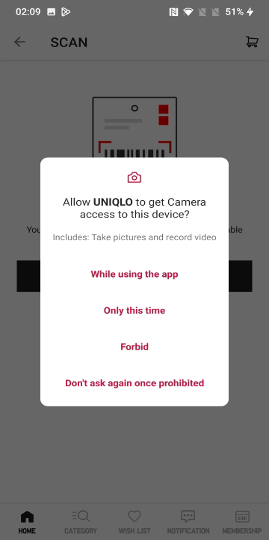}
  \caption{Invoke-time}
  \label{fig:sfig2}
\end{subfigure}
\begin{subfigure}{.315\linewidth}
  \centering
  \includegraphics[width=.98\linewidth]{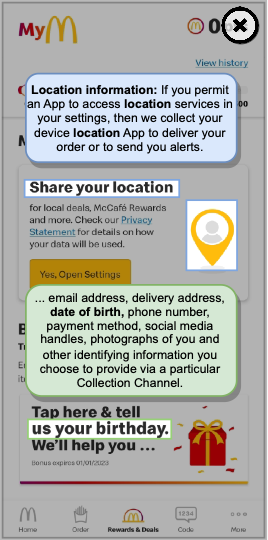}
  \caption{Context-aware}
  \label{fig:sfig2}
\end{subfigure}
\vspace{-5pt}
\caption{
The leftmost is the install-time reminder of required permissions~\cite{intro_installtime}; the middle is the reminder that pops up the first time to invoke permission; and the rightmost is our contextual privacy policy for mobile apps.
}
\label{fig_introduction}
\vspace{-5pt}
\end{figure}
%

To prevent privacy violations, the U.S. Department of Health, Education, and Welfare proposed a set of Fair Information Practice Principles in 1973~\cite{FIPP1973, landesberg1998privacy}. Later on, the Federal Trade Commission~\cite{obar2020biggest} adopted these principles as ``\textit{the notice and choice privacy framework,}'' which is commonly regarded as the foundation for modern privacy regulations such as European General Data Protection Regulation (GDPR)~\cite{GDPR}, California Consumer Privacy Act (CCPA)~\cite{CCPA}, and Australian Privacy Principles (APP)~\cite{APP}. Today, privacy policies have become the most critical approach to safeguarding individuals' privacy and digital security~\cite{caramujo2015analyzing, kemp2020concealed, perez2018review}.
Despite the provision of clear, concise, and transparent privacy policies, users often grant permissions to mobile apps to access their personal information without comprehensively understanding these policies.
This phenomenon persists even when the policies are readily available for review. Obar and Oeldorf-Hirsch~\cite{obar2020biggest} have demonstrated that a significant portion of users, approximately 74\%, opt for the ``quick join'' clickwrap option without engaging with the privacy policies, primarily due to the lengthy and detailed nature of these documents. The findings of a study conducted on 75 prominent mobile applications and websites~\cite{Blakkarly2022privacy} further support this observation, revealing that the average length of privacy policies is around 4,000 words, requiring an estimated 16 minutes to read. These findings indicate that privacy policies have become increasingly challenging to read and understand~\cite{amos2021privacy}.

Many studies have been conducted to improve the presentation and readability of privacy policies from various perspectives. For example, Kelly et al.~\cite{kelley2009nutrition, kelley2010standardizing, kelley2013privacy} suggested a concise and organized approach, namely privacy nutrition labels (PNL), to help customers better understand how their information is gathered and utilized. Moreover, PNL effectively facilitates their comprehension of software applications' privacy practices~\cite{ciocchetti2008future, cranor2012necessary, pan2023toward}. Cranor~\cite{cranor2002web} proposed a protocol, i.e., Platform for Privacy Preferences, that allows website organizations to declare their intended use of information collected from website users. Another research direction focuses on privacy icons~\cite{holtz2011towards, holtz2011privacy, efroni2019privacy}, small graphic symbols representing a privacy concept or practice, to assist users in understanding data handling practices by providing a visual shorthand for complex privacy policies. These standardized privacy icons across multiple services and platforms also help reduce confusion and make privacy policies more accessible to users~\cite{holtz2011privacy}.

A recent development in the evolution of privacy policies is ``just-in-time'' privacy notices. Specifically, these notices offer users precise information about data collection at the moment of interaction. The U.K. GDPR has included the just-in-time notice as a recommended practice in its code of conduct for creating precise privacy policies~\cite{ICOjustintime}. However, most of the existing just-in-time notices focus on when users install applications. For example, an ``install-time'' permission reminder may pop up when a user installs a new Android app, clarifying what permissions will be granted and how they will be utilized (see Fig.~\ref{fig_introduction}a). Another example is shown in Fig.~\ref{fig_introduction}b, which is a camera access reminder triggered at the first ``invoke-time''.
Some studies~\cite{felt2010effectiveness, ma2014android} have criticized the effectiveness of existing ``install-time'' privacy reminders since security warnings for dangerous permissions are often decoupled from the context, failing to raise awareness among users~\cite{felt2010effectiveness}. 
To mitigate this problem, researchers have proposed a \textit{contextual privacy policy} (CPP) to offer immediate privacy notices in their corresponding contexts~\cite{Feth2017, ortloff2020implementation, onu2020contextual, windl2022automating}. Contextual privacy policies aim to break down policies into shorter snippets and only display them in their corresponding contexts, which greatly helps users to comprehend obscure language and reduces the amount of text that must be read at once.
Privacy-related contexts are areas where data practices apply to GUIs.
Fig.~\ref{fig_introduction}c shows an example of our contextual privacy policy on mobile. Privacy-related contexts on the screen are detected out, and their corresponding (in the same color) privacy policy segments are retrieved and displayed, with emboldened key information.
By doing so, privacy notices are closely and timely incorporated with the context, effectively raising users' awareness and delivering privacy information.

Compared to other software applications, mobile apps have more sophisticated designs, compact Graphical User Interfaces (GUIs), and frequent human-device interactions; hence, we have to identify the contexts that need to display CPPs in the mobile apps. We first conduct a comprehensive survey to collect and identify data types that have been covered in regulation principles, industry standards, and previous academic studies about privacy practices on mobile apps (See Section~\ref{sec_background_survey}). 
Specifically, we identify 12 CPP-related data types in mobile apps and collect their keyword lists from surveyed literature. We then accumulate a full list of CPP-related GUI components for nine data types from RICO-icon, one of the largest mobile GUI icon datasets. Based on this information, we establish the benchmark dataset named \textbf{\textsc{Cpp4App}}, containing over 600 privacy-related contexts with their corresponding privacy policy segments in the mobile scenario.
Then we propose a novel multi-modal framework, namely \textbf{\textsc{SeePrivacy}}, that can automatically generate contextual privacy policies for mobile apps. Specifically, we utilize Computer Vision (CV) techniques and a pre-trained Large Language Model to understand the mobile GUI screenshots, detecting privacy-related contexts. In addition, we employ Natural Language Processing (NLP) techniques to analyze the privacy policy documents, fetching the corresponding policy segments for the contexts.
Our framework yields 0.76, 0.86, and 0.87 in terms of accuracy, precision, and recall, respectively, on category-wise context identification, and an overall 83.6\% coverage rate for privacy-related contexts detection as per the screenshot.
As for extracting corresponding segments, our framework achieves 0.92 accuracy, 0.94 precision, and 0.98 recall. Also, 96\% of retrieved policy segments can be correctly matched with their contexts.
In addition, we conduct a usability evaluation to further validate its functionality and usefulness in practice. More than 77\% of participants do not disagree that the retrieved privacy policy segment explains the detected context, which is underlined by a 3.75 mean and 4 median on a 5-point Likert scale.
Overall, \textsc{SeePrivacy} receives 4.53 out of 5 from users about its usefulness. Moreover, the proposed framework does not require source code or additional APIs; hence, developers will easily deploy the framework with lower security concerns. Moreover, the generated CPPs offer concise, transparent, and timely privacy information that assists users in making choices about whether they should interact with GUI components. The main contributions of our paper are as follows:
\begin{itemize} [leftmargin=*]
\item To the best of our knowledge, we are the first to propose a novel framework, i.e., \textsc{SeePrivacy}\footnote{The demonstration website: \url{https://cpp4app.github.io/}}, to automatically generate contextual privacy policies for mobile apps.
\item We build the benchmark dataset, i.e., \textsc{Cpp4App}, for contextual privacy policies on mobile apps and evaluate \textsc{SeePrivacy} on it to demonstrate the effectiveness of our approach. 
\item We conduct a user study to further validate \textsc{SeePrivacy}'s functionality and usability in practice.

\end{itemize}

\begin{table*}[!t]
  \caption{Data types' existence in literature.}
  \vspace{-5pt}
  \label{tab_survey}
  \centering
  \resizebox{.95\textwidth}{!}{%
  \begin{tabular}{l|cccccc|cccccc}
    \hline
    \rowcolor{lightgray!15}
    &                                                                                        \multicolumn{6}{c|}{Basic Personal Identifiable Information}                                                                       & \multicolumn{6}{c}{Other Personal Information}                                                                                     \\ 
    \hline
    \rowcolor{lightgray!50}
    &\textbf{Name} & \textbf{Birthday} & \textbf{Address} & \textbf{Phone} & \textbf{Email} &\textbf{Profile}
    & \textbf{Contacts} & \textbf{Location} & \textbf{Photos} & \textbf{Voices} & \textbf{Financial Info} 
    & \textbf{Social Media} \\
    \hline
    
    \rowcolor{lightgray!85}
    \textbf{Administration}
    & & & & & & & & & & & &\\
    \rowcolor{lightgray!15}
    GDPR~\cite{GDPR} 
    & \CIRCLE &\Circle &\CIRCLE &\CIRCLE &\CIRCLE &\CIRCLE
    & \Circle &\Circle &\Circle &\Circle &\Circle &\Circle 
    \\
    \rowcolor{lightgray!50}
    CCPA~\cite{CCPA} 
    & \CIRCLE &\Circle &\Circle &\Circle &\CIRCLE &\CIRCLE
    & \Circle &\CIRCLE &\Circle &\Circle &\CIRCLE &\Circle 
     \\
    \rowcolor{lightgray!15}
    CalOPPA~\cite{CalOPPA} 
    & \CIRCLE &\Circle &\CIRCLE &\CIRCLE &\CIRCLE &\Circle
    & \Circle &\CIRCLE &\Circle &\Circle &\CIRCLE&\CIRCLE 
    \\
    \rowcolor{lightgray!50}
    COPPA~\cite{COPPA} 
    & \CIRCLE &\Circle &\CIRCLE &\CIRCLE &\CIRCLE &\CIRCLE
    & \Circle &\CIRCLE &\CIRCLE &\CIRCLE &\Circle&\Circle 
    \\
    
    \rowcolor{lightgray!15}
    APP~\cite{APP} 
    &\CIRCLE &\Circle &\CIRCLE &\CIRCLE &\CIRCLE &\CIRCLE
    &\Circle &\CIRCLE &\CIRCLE &\Circle &\CIRCLE &\Circle
    \\
    
    \rowcolor{lightgray!85}
    \textbf{Industry}
    & & & & & &  & & & & & & \\
    \rowcolor{lightgray!15}
    Google Play~\cite{googledatasafety} 
    & \CIRCLE & \CIRCLE & \CIRCLE & \CIRCLE & \CIRCLE & \CIRCLE
    & \CIRCLE & \CIRCLE & \CIRCLE & \CIRCLE & \CIRCLE & \CIRCLE
    \\
    \rowcolor{lightgray!50}
    Apple App~\cite{appleprivacy} 
    & \CIRCLE & \Circle & \CIRCLE & \CIRCLE & \CIRCLE & \CIRCLE
    & \CIRCLE & \CIRCLE & \CIRCLE & \CIRCLE & \CIRCLE & \CIRCLE
    \\
    \rowcolor{lightgray!15}
    Huawei AppGallery~\cite{Huaweiprivacy} 
    & \CIRCLE & \CIRCLE & \CIRCLE & \CIRCLE & \CIRCLE & \CIRCLE
    & \CIRCLE & \CIRCLE & \CIRCLE & \CIRCLE & \CIRCLE & \CIRCLE 
    \\
    \rowcolor{lightgray!50}
    Amazon Appstore~\cite{Amazonprivacy} 
    & \Circle & \Circle & \Circle & \CIRCLE & \CIRCLE & \Circle
    & \CIRCLE & \CIRCLE & \CIRCLE & \CIRCLE & \CIRCLE & \CIRCLE
    \\
    \rowcolor{lightgray!15}
    Samsung Galaxy Store~\cite{samsungprivacy}
    & \Circle & \Circle & \Circle & \CIRCLE & \Circle & \Circle 
    & \CIRCLE & \CIRCLE & \CIRCLE & \CIRCLE & \Circle & \Circle
    \\
    
    \rowcolor{lightgray!85}
    \textbf{Academia} &&&&&& &&&&&&\\
    \rowcolor{lightgray!15}
    Yu et al.~\cite{yu2015autoppg} 
    & \CIRCLE & \CIRCLE & \CIRCLE & \CIRCLE & \CIRCLE & \CIRCLE
    & \CIRCLE & \CIRCLE & \CIRCLE & \CIRCLE & \Circle & \Circle
    \\
    \rowcolor{lightgray!50}
    Wilson et al.~\cite{wilson2016creation} 
    & \CIRCLE & \CIRCLE & \CIRCLE & \CIRCLE & \CIRCLE & \CIRCLE
    & \CIRCLE & \CIRCLE & \Circle & \Circle & \CIRCLE & \CIRCLE
    \\
    \rowcolor{lightgray!15}
    Harkous et al.~\cite{harkous2018polisis} 
    & \CIRCLE & \CIRCLE & \CIRCLE & \CIRCLE & \CIRCLE & \CIRCLE
    & \CIRCLE & \CIRCLE & \Circle & \Circle & \CIRCLE & \CIRCLE
    \\
    \rowcolor{lightgray!50}
    Kaur et al.~\cite{kaur2018comprehensive} 
    & \CIRCLE & \CIRCLE & \CIRCLE & \CIRCLE & \CIRCLE & \CIRCLE
    & \Circle & \CIRCLE & \Circle & \Circle & \CIRCLE  & \Circle
    \\
    \rowcolor{lightgray!15}
    Yu et al.~\cite{yu2018enhancing}
    & \Circle & \Circle & \Circle & \Circle & \Circle & \CIRCLE
    & \CIRCLE & \CIRCLE & \CIRCLE & \CIRCLE & \Circle & \Circle
    \\
    \rowcolor{lightgray!50}
    Zimmeck et al.~\cite{zimmeck2019maps} 
    & \Circle & \Circle & \Circle & \CIRCLE & \CIRCLE & \CIRCLE
    & \Circle & \CIRCLE & \Circle & \Circle & \Circle & \CIRCLE
    \\
    \rowcolor{lightgray!15}
    Zhang et al.~\cite{zhang2020does} 
    & \CIRCLE & \Circle & \CIRCLE & \CIRCLE & \CIRCLE & \Circle
    & \Circle & \CIRCLE & \Circle & \Circle & \Circle & \Circle
    \\
    \rowcolor{lightgray!50}
    Sun and Xue~\cite{sun2020quality} 
    & \CIRCLE & \Circle & \Circle & \CIRCLE & \CIRCLE & \Circle
    & \Circle & \CIRCLE & \Circle & \Circle & \CIRCLE & \Circle
    \\
   
    \rowcolor{lightgray!15}
    Li et al.~\cite{li2021developers} 
    & \CIRCLE & \Circle & \CIRCLE & \CIRCLE & \CIRCLE & \CIRCLE
    & \Circle & \CIRCLE & \CIRCLE & \CIRCLE & \Circle & \Circle
    \\
    \rowcolor{lightgray!50}
    Zimmeck et al.~\cite{zimmeck2021privacyflash} 
    & \Circle & \Circle & \Circle & \Circle & \Circle & \Circle
    & \Circle & \CIRCLE & \CIRCLE & \CIRCLE & \Circle & \CIRCLE
    \\
     
    \rowcolor{lightgray!15}
    Xie et al.~\cite{xie2022scrutinizing} 
    & \CIRCLE & \CIRCLE & \CIRCLE & \CIRCLE & \CIRCLE & \CIRCLE
    & \CIRCLE & \CIRCLE & \Circle & \Circle & \CIRCLE & \Circle
    \\
  \hline
\end{tabular}
}%
\vspace{-15pt}
\end{table*}

\section{Background and Privacy-related Context Identification for Mobile Apps}
\label{sec_background}


The concept of ``\textit{contextual privacy policies}'' (CPPs) was first proposed in 2004. Specifically, Bolchini et al.~\cite{bolchini2004need} identified that reading privacy policies is a monolithic block when users attempt to retrieve privacy information; thus, they invented a systematic approach to restructuring the privacy policies based on user interaction contexts. Later on, Feth~\cite{Feth2017} introduced the idea of ``\textit{contextual privacy statements.}'' Rather than using a ``one-size-fits-all'' privacy policy that must suit every usage scenario, these contextual privacy statements offer precise details about privacy and data protection in a specific activity or use case. Recently, Windl et al.~\cite{windl2022automating} proposed the design and architecture of PrivacyInjector, a production AI tool that can automatically generate contextual privacy policies for websites. CPPs have been demonstrated to effectively help users understand the privacy practices and consequences of potential privacy risks~\cite{ortloff2020implementation}. CPPs also help increase privacy awareness and transparency for users~\cite{Al2012Awareness, 7927931, masotina2022transparency}.


Compared to websites, mobile apps often collect personal information in multiple ways through device sensors, user inputs, and third-party integration~\cite{pan2023large}; however, disclosing how users' personal information is gathered, utilized, and disseminated is a crucial problem. As the size of mobile apps is enormous and is still rapidly growing, a framework to automatically generate contextual privacy policies for mobile apps is urgently needed. 
In this work, we aim to propose a framework that can automatically generate CPPs, but first, we need to have clear definitions of privacy-related context in the mobile scenario.

\vspace{-3pt}
\subsection{Privacy-related Context Identification for Mobile Apps}
\label{sec_background_survey}

Privacy-related contexts are considered areas on the GUIs where data practices are applied~\cite{windl2022automating}. To specify the data practices for this work, we propose three questions:
\begin{itemize}[leftmargin=*]
\item What data practices should we include?
\item What data practices are of concern and desire?
\item What information is delivered by common privacy policies?
\end{itemize}
%
Note that we have not found any studies that discuss contextual privacy policies for mobile apps. Specifically, most research only focuses on improving the presentation and readability of privacy policies~\cite{kelley2009nutrition, kelley2013privacy, kelley2010standardizing, cranor2002web, ciocchetti2008future}. In this paper, we conduct a comprehensive survey (see Table~\ref{tab_survey}) on the data practices identified and discussed by three groups: administration, market standards from industry, and academic research papers, to answer the three questions. 


\emph{Administration.} We often see the administration as the regulator of the ecosystem. According to the United Nations Conference on Trade and Development, 137 out of 194 countries and regions have launched data protection and privacy legislation~\cite{UNCTAD}. These requirements outline data practices that should be included in privacy notices. We select five privacy regulations, such as GDPR~\cite{GDPR}, CCPA~\cite{CCPA}, CalOPPA~\cite{CalOPPA}, COPPA~\cite{COPPA}, and APP~\cite{APP}, to understand the data practices from the administration's point of view. If users notice a specific data type, we denote its existence.


\emph{Industry.} Mobile app platforms commonly have specific privacy notice requirements for curated apps in the market, especially for the involved data types. For example, the Google Play app store defined 14 main data categories and 39 sub-categories for their data safety reports that all apps need to claim\cite{googledatasafety}. Those standards have been developed through extensive user studies and market testing. Therefore, market standards can reflect the users' concerns and expectations in practice.

\emph{Academia.}
%
Researchers have studied the privacy policies for software applications from various perspectives, including content categorization and extraction~\cite{wilson2016creation, kaur2018comprehensive, zimmeck2019maps, li2021developers}, automated generation~\cite{yu2015autoppg, sun2020quality, zimmeck2021privacyflash}, and compliance analysis~\cite{yu2018enhancing, zimmeck2019maps, zhang2020does, xie2022scrutinizing}. As there are differences among privacy policies, corpus selections, and tasks among numerous problems, their data types vary as well. By surveying the academic research literature, we summarize the data types covered by common privacy policies. 

\begin{table*}[!t]

\caption{The identification of privacy-related contexts.}
\vspace{-5pt}
  \label{tab_keyword_list}
\centering

  \resizebox{.95\textwidth}{!}{%
\begin{tabular}{|l|l|l|l|c|}
\hline
\textbf{Data types} & \textbf{Description} & \textbf{Keywords list}  & \textbf{Related RICO-icon class} & \textbf{Icon examples}\\ 
\hline
\hline
Name & How a user refers to themselves & \makecell[l]{name, first name, last name, full name, real name, surname,\\ family name, given name}  & n.a. &  
\raisebox{-0.0\totalheight}{%
n.a.
\begin{minipage}{.04\textwidth}
  \includegraphics[width=0.85\linewidth]{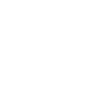}
\end{minipage}
}%
\\ \hline
Birthday & A user’s birthday & birthday, date of birth, birth date, DOB, birth year & n.a. & 
\raisebox{-0.0\totalheight}{%
n.a.
\begin{minipage}{.04\textwidth}
  \includegraphics[width=0.85\linewidth]{Figures/placeholder.png}
\end{minipage}
}%
\\ \hline
Address & A user’s address  & \makecell[l]{mailing address, physical address, postal address, billing address, \\shipping address, residential address, residence, personal address} &n.a& 
\raisebox{-0\totalheight}{%
n.a
\begin{minipage}{.04\textwidth}
  \includegraphics[width=0.85\linewidth]{Figures/placeholder.png}
\end{minipage}
}%
\\ \hline
Phone & A user’s phone number & \makecell[l]{hone, phone number, mobile, mobile phone, mobile number, \\telephone, telephone number, call}  & [43] Call &
\raisebox{-0.06\totalheight}{%
\begin{minipage}[c]{.04\textwidth}
  \includegraphics[width=0.92\linewidth]{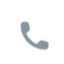}
\end{minipage}
}%
 \raisebox{-0.06\totalheight}{%
\begin{minipage}[c]{.04\textwidth}
  \includegraphics[width=0.8\linewidth]{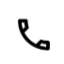} 
\end{minipage}
}%
\\ \hline
Email & A user’s email address  
 \raisebox{-0\totalheight}{%
\begin{minipage}{.04\textwidth}
  \includegraphics[width=0.85\linewidth]{Figures/placeholder.png}
\end{minipage}
}%
& email, e-mail, email address, e-mail address  & [6] Email & 
\raisebox{-0.06\totalheight}{%
\begin{minipage}[c]{.04\textwidth}
  \includegraphics[width=0.85\linewidth]{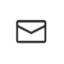}
\end{minipage}
}%
 \raisebox{-0.05\totalheight}{%
\begin{minipage}[c]{.04\textwidth}
  \includegraphics[width=0.8\linewidth]{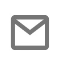} 
\end{minipage}
}%
\\ \hline
Profile & A user’s account information 
 \raisebox{-0\totalheight}{%
\begin{minipage}{.04\textwidth}
  \includegraphics[width=0.8\linewidth]{Figures/placeholder.png}
\end{minipage}
}%
& profile, account
& [49] Avatar & 
\raisebox{-0.06\totalheight}{%
\begin{minipage}[c]{.04\textwidth}
  \includegraphics[width=0.85\linewidth]{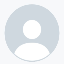}
\end{minipage}
}%
 \raisebox{-0.05\totalheight}{%
\begin{minipage}[c]{.04\textwidth}
  \includegraphics[width=0.8\linewidth]{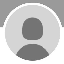} 
\end{minipage}
}%
\\ \hline
\hline
Contacts &  \makecell[l]{A user’s contact information, or the \\access to the contact permission} 
& contacts, phone-book, phone book, device's address book  &[68] Group, [3] Follow &
\raisebox{-0.05\totalheight}{%
\begin{minipage}[c]{.04\textwidth}
  \includegraphics[width=0.8\linewidth]{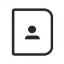}
\end{minipage}
}%
 \raisebox{-0.05\totalheight}{%
\begin{minipage}[c]{.04\textwidth}
  \includegraphics[width=0.8\linewidth]{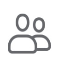} 
\end{minipage}
}%
\\ \hline
Location &  \makecell[l]{A user’s location information, or the \\access to the location permission} & \makecell[l]{location, locate, 
geography,\\ geo, geo-location, precision location} & \makecell[l]{[40] Location crosshair, \\  \text{[72] Location}} &
\raisebox{-0.03\totalheight}{%
\begin{minipage}[c]{.04\textwidth}
  \includegraphics[width=0.8\linewidth]{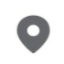}
\end{minipage}
}%
\raisebox{-0.05\totalheight}{%
\begin{minipage}[c]{.04\textwidth}
  \includegraphics[width=0.8\linewidth]{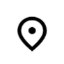}
\end{minipage}
}%
\\ \hline
Photos & \makecell[l]{A user’s photos, videos, or the access \\to the camera permission} & \makecell[l]{camera, photo, scan, album, picture, gallery, photo library,\\ storage, image, video, scanner, photograph} & \makecell[l]{[42] Photo, [56] Videocam,\\ \text{[82] Wallpaper}}  &
\raisebox{-0.05\totalheight}{%
\begin{minipage}[c]{.04\textwidth}
  \includegraphics[width=0.8\linewidth]{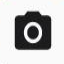}
\end{minipage}
}%
\raisebox{-0.05\totalheight}{%
\begin{minipage}[c]{.04\textwidth}
  \includegraphics[width=0.8\linewidth]{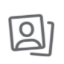}
\end{minipage}
}%
\\ \hline
Voices&  \makecell[l]{A user’s voices, recordings, or the access\\ to the microphone permission} & microphone, voice, mic, speech, talk  & [91] Microphone  & 
\raisebox{-0.05\totalheight}{%
\begin{minipage}[c]{.04\textwidth}
  \includegraphics[width=0.75\linewidth]{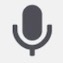}
\end{minipage}
}%
\raisebox{-0.05\totalheight}{%
\begin{minipage}[c]{.04\textwidth}
  \includegraphics[width=0.75\linewidth]{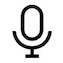}
\end{minipage}
}%
\\ \hline
Financial info & \makecell[l]{Information about a user’s financial \\accounts, purchases, or transactions}  & \makecell[l]{credit card, company, companies, organization,\\ organizations, pay, payment}  & [61] Cart  &
\raisebox{-0.05\totalheight}{%
\begin{minipage}[c]{.04\textwidth}
  \includegraphics[width=0.85\linewidth]{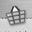}
\end{minipage}
}%
\raisebox{-0.05\totalheight}{%
\begin{minipage}[c]{.04\textwidth}
  \includegraphics[width=0.8\linewidth]{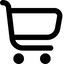}
\end{minipage}
}%
\\ \hline
Social media &  \makecell[l]{A user's social media information, or \\the access to social media accounts}
& social media, Facebook, Twitter, socialmedia, share  & [77] Facebook, [89] Twitter &
\raisebox{-0.05\totalheight}{%
\resizebox{.042\textwidth}{!}{%
\begin{minipage}[c]{.041\textwidth}
  \includegraphics[width=0.8\linewidth]{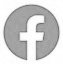}
\end{minipage}
}%
}%
\raisebox{-0.05\totalheight}{%
\resizebox{.04\textwidth}{!}{%
\begin{minipage}[c]{.04\textwidth}
  \includegraphics[width=0.8\linewidth]{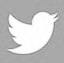}
\end{minipage}
}%
}%
 \\ \hline
\end{tabular}
}
\vspace{-5pt}
\end{table*}


\vspace{-5pt}
\subsection{Multi-modal Privacy-related Contexts}
\label{sec_background_context}

Privacy-related contexts encompass the areas within GUIs where data practices could be applicable.
In this paper, we identify two types of GUI components that may collect personal information through human-device interactions: literal texts and graphical icons.
On one hand, literal texts are often considered the most fundamental information sources in mobile app GUIs.
On the other hand, GUIs are specifically designed to convey essential information as efficiently as possible.
More graphical icons are employed to replace literal texts, as these icons are concise and provide users with the necessary information at a glance, significantly reducing the effort to read the literal texts~\cite{xiao2019iconintent}. 

\emph{Keyword collection.} 
After reviewing literal texts describing data types, we build a keyword list for contextual privacy policies. Specifically, we collect keywords from the academic research literature. We exclude ambiguous words, such as ``address'' since they may represent residential addresses or email addresses.
To improve unification and consistency, the list is made case-insensitive. 
The complete keyword list is shown in Table~\ref{tab_keyword_list}.

\emph{Icon collection.}
We notice that icons often appear on the app interface as representatives of a specific type of privacy information. For instance, the pattern of the camera usually relates to the acquisition of the camera's permission by mobile apps. 
Liu et al.~\cite{Liu:2018:LDS:3242587.3242650} proposed an icon taxonomy with 99 common classes. They also provided one of the largest icon datasets, with over 118K icons. We then semantically match their RICO-icon classes with the data types identified from the survey (See Table~\ref{tab_keyword_list}).

\section{Cpp4App: The Benchmark Dataset}
\label{sec_dataset}

%
\begin{figure*}[t!]
  \centering
  \includegraphics[width=.9\linewidth]{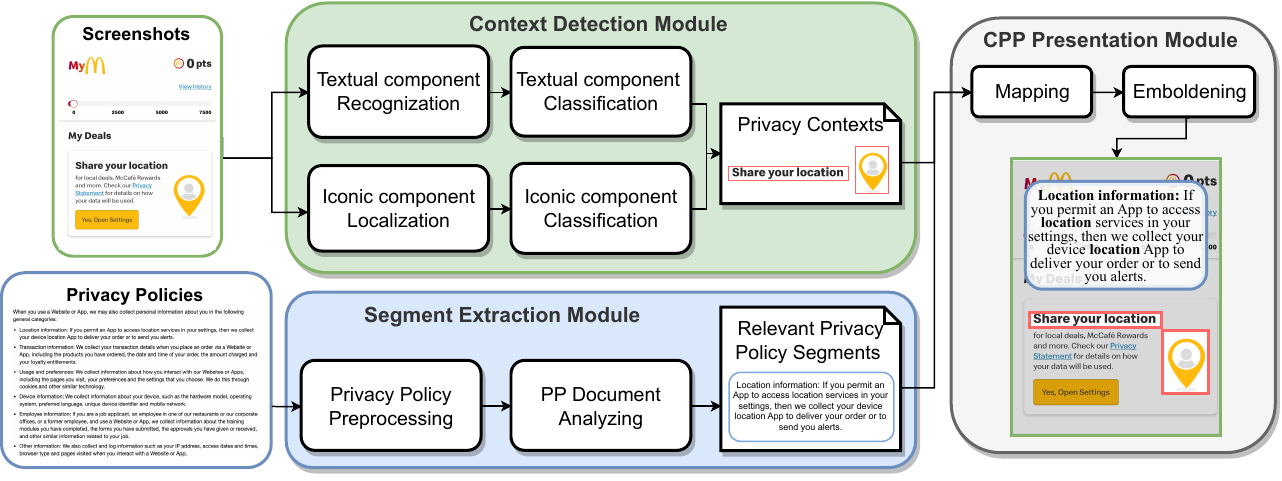}

  \caption{An overview of \textsc{SeePrivacy}.}
  \label{fig_pipeline}
 \vspace{-5pt}
\end{figure*}
%

There are some existing datasets for mobile GUI understanding~\cite{leiva2020understanding, bunian2021vins, deka2017rico}; however, we realize two main challenges in the current mobile app datasets that drive us to collect a new dataset specifically for automatically generating contextual privacy policies:
\begin{itemize}[leftmargin=*]
\item \textbf{Temporal inconsistency between GUIs and privacy policies.} 
As the privacy practices of mobile apps are constantly changing, their corresponding privacy policies also need to be updated. Moreover, the GUIs of mobile apps also evolve rapidly. For these reasons, it is crucial to maintain consistency between screenshots (GUIs) and their temporally corresponding privacy policies to ensure accuracy and reliability. In current GUI datasets, such as Rico~\cite{deka2017rico}, most privacy policies for mobile apps are either outdated or not publicly available.

\item \textbf{No annotations about privacy-related contexts in GUIs.}
As far as we know, we are the first to propose a framework that automatically generates contextual privacy policies for mobile apps. To evaluate the performance of the proposed framework, we expect benchmark datasets to contain two features, i.e., the bounding boxes of privacy-related contexts on the GUI screenshots and the mapping relations between privacy-related contexts and privacy policy segments. However, there are no existing benchmark datasets that include these two features. 
\end{itemize}
In this work, we need to construct a benchmark dataset to accurately evaluate the effectiveness of generating contextual privacy policies. We present steps to create the benchmark dataset in the following subsections. 

\subsection{Mobile Apps Selection}

Android mobile apps are often transparent and favorable to academic research~\cite{li2019rebooting, wei2017survey}. There are numerous app markets accessible for the Android platform; however, the Google Play app store is the most extensive and easily accessible, offering over two million apps~\cite{malavolta2015end, wang2018beyond}. Moreover, this store is one of the leading stores requiring mobile apps to provide a privacy policy link on their homepages~\cite{googledeveloper}. Therefore, users can directly obtain privacy policy links when downloading Android mobile apps from the Google Play app store.


According to previous studies, popular mobile apps are more likely to have rigorous privacy practices~\cite{zimmeck2019maps, liu2021have, USENIX_2022_GEODIFF}. We employ information from AppBrain~\cite{appbrain}, one of the largest third-party platforms monitoring app statistics for the Google Play app store, to identify the most popular 200 mobile apps. We follow three strategies to collect mobile apps for our paper. First, the Google Play app store assigns a specific category to each mobile app. Moreover, mobile apps from various categories vary in GUIs, functions, and privacy practices. To ensure the diversity of mobile apps, we collect the most popular apps across 14 distinct categories. Second, we exclude mobile apps that have privacy policies smaller than 200 words or smaller than 2 KB, as they may contain invalid privacy policies~\cite{USENIX_2022_GEODIFF, liu2023appcorp}. Lastly, we filter out mobile apps with non-English privacy policies. 
The selection criteria drive 30 candidates' mobile apps for following fine-grained annotation.


\vspace{-2pt}
\subsection{Privacy Policy Segment Annotation Strategy}
%
For each data type, we ask our annotators to read the privacy policies and select relevant segments for each privacy-related context. To guarantee the quality of the annotation process, we set a maximum reading speed of 250 words per minute for annotators. If the annotator is unable to identify any relevant segments from privacy policies, then it will be labeled as ``No relative information is found in the privacy policy''.

\subsection{Privacy-related Context Annotation Strategy}
Based on the analysis of mobile app GUI compositions~\cite{10essentialscreens, 9mainscreens}, we collect specific pages of each mobile app to ensure its essential privacy-related GUIs are included in our dataset. Specifically, the pages are listed as follows: 
\begin{itemize}[leftmargin=*]
\item \emph{Registration/Login page.} 
Mobile apps often collect user registration information on this page to create a new account, which may include personal and other private information. Certain mobile apps mandate new users to register before accessing their features, making it impossible for these users to bypass the registration or login page.

\item \emph{Homepage.} This is the first page that appears after entering a mobile app. It contains essential features and information related to its app's functionality, which may require you to provide personal information or grant device permissions. Additionally, we also gather screenshots of the main pages next to the homepage.

\item \emph{Profile page.} This page may store the personal information of users. For example, some mobile apps may request permission to access the camera or photo roll to provide customization options for the account's public appearance. 

\item \emph{Setting page.} This page may  involve privacy-related customization, such as sensor permissions. 

\item \emph{Onboarding page.} This page is considered a tutorial or introduction page that presents the app's features and benefits. 

\item \emph{Map page.} We visit this page in various categories of mobile apps, such as travel, posting, food delivery, and shopping. These mobile apps tend to collect users' location information on this page.

\item \emph{Essential pages for specific categories of apps.} These include the catalog page and product card page in shopping apps, the post editing page in forum apps, and the feed page in social media apps.
\end{itemize}


For each screenshot of its page, annotators need to manually read all the texts and check whether they contain any word in our CPP-related keyword list mentioned in Table~\ref{tab_keyword_list}. If the text contains CPP-related keywords, it will be labelled as the corresponding data type and marked with a bounding box. Then, annotators need to detect all icons displayed on the screenshot if the icon belongs to a privacy-related data type. If so, the icon will be labelled as the corresponding class and indicated with a bounding box. We employ the \href{https://pypi.org/project/opencv-python/}{\textit{OpenCV-Python}} library to draw all bounding boxes.

\begin{table}[!t]
\centering
  \caption{The statistical analysis of benchmark dataset.}
  \label{tab_dataset_summary}
  \begin{tabular}{l | c}
    No. Mobile applications \& Privacy policies & 30  \\
    \hline
    \hline
    No. Screenshots & 230 \\
    Screenshots per Mobile application & 7.67\\
    No. Privacy-related contexts & 644\\
    Contexts per Mobile application & 21.5\\
    \hline
    \hline
    No. Words in Privacy policies & 141,292 \\
    Words per Privacy policy & 4,709\\
\end{tabular}
\end{table}

\vspace{-5pt}
\subsection{Setups and Statistics}
\label{dataset_setup}
To guarantee the consistency between GUIs and privacy policies during this work, an Android mobile phone, i.e., the OnePlus 7 Pro, Android 11, and Hydrogen OS 11.0.9.1.GM21, is employed to download all the selected mobile apps. Their corresponding privacy policies are downloaded simultaneously. Then the auto-update features in the Google Play app store and mobile apps are turned off. Also, the auto-update over Wi-Fi feature is disabled in the system's general settings on this mobile phone. 


To select pages of each mobile app, two annotators explore the mobile app and take screenshots of these pages. In total, we have 230 screenshots from 30 mobile apps. For each screenshot, the two annotators label privacy-related contexts and their corresponding privacy policy segments independently. If the annotators have any labeling conflicts, they discuss them and reach a consensus. In total, we have 644 labeled privacy-related contexts, with each context corresponding to a privacy policy segment, and we name this benchmark dataset as \textsc{Cpp4App}. The statistical analysis is shown in Table~\ref{tab_dataset_summary}.


\section{SeePrivacy: The Framework}


In this section, we present our multi-modal framework, namely \textsc{SeePrivacy}, that automatically generates contextual privacy policies (CPPs) for given mobile apps (see Fig.~\ref{fig_pipeline}). The framework takes privacy policies and screenshots from a given mobile app as inputs. These inputs are fed into three main modules to automatically construct CPPs for the mobile app. We list the three main modules of our framework as follows: 

\begin{itemize}[leftmargin=*]
\item \textit{Context Detection Module:} This module aims to locate contexts in which we should display contextual privacy policies from a given mobile app's screenshot. 
\item \textit{Segment Extraction Module:} This module aims to extract relevant privacy policy segments from a given mobile app's privacy policy.
\item \textit{CPPs Presentation Module:} This module maps privacy policy segments to their corresponding context, rendering the key information, to eventually generate contextual privacy policies for mobile apps. 
\end{itemize}
Details of each module are explained in the following subsections. 

\subsection{Context Detection Module}


Our framework only takes a given screenshot to detect privacy-related contexts without obtaining the source code or extra run-time APIs, therefore our framework is easier to deploy and has high adaptability. We choose visual components instead of widgets as atomic elements in our GUI understanding since some informative widgets are too coarse-grained to be a context.
To localize the visual components on the screenshot for potentially indicating or suggesting personal information collection, we propose a divide-and-conquer method that can detect texts and icons separately and identify the contexts together.

\textbf{Textual GUI components.} 
%
To detect and localize privacy-related texts from a given screenshot, we apply a state-of-the-art ultra-lightweight OCR system, namely PaddlePaddle-OCR (PP-OCR)~\cite{li2022pp}, for the mobile scenario. After obtaining these texts, we employ a pre-trained Large Language Model, i.e., GPT-3.5, to classify their data types (see Table~\ref{tab_keyword_list}). Specifically, we first inject the data type list and their descriptions displayed in Table~\ref{tab_keyword_list}, and then for every \textit{\{detected text\}}, we propose the following prompt: 

\begin{mdframed}[roundcorner=8pt]
Question: Is this piece of text ``\textit{\{detected text\}}'' related to any following privacy information data types? Or not relevant to any of them? ONLY answer the data type or "not relevant". ONLY use the provided data type list. 

\noindent $\quad$
\vspace{-7pt}

\noindent Answer:
\end{mdframed}
If the answer contains a data type, we will regard the \textit{\{detected text\}} as relevant to this data type; otherwise, it does not belong to any data type.

\textbf{Iconic GUI components - Localization.} 
To localize iconic GUI components from their screenshots, we follow previous work~\cite{xie2020uied} and propose a new set of detection rules to filter out irrelevant objects. Specifically, we have four rules presented as follows: (a) Filtering out elements whose area is greater than 10\% of the total area of the screenshot, aiming to remove the bounding boxes and large images; (b) Filtering out elements whose area is less than 5\% of the total area of the screenshot, aiming to remove the small noises; (c) Removing non-squarish elements whose aspect ratio (width to height) is less than 0.6, ruling out control bars and top/bottom banners; (d) Removing elements that area is overlapped with OCR results since some letters in large font-size are also detected as potential icons. 
By applying these rules to our dataset, we can obtain the iconic GUI components' position on their screenshots.

\textbf{Iconic GUI components - Classification.}
%
CNN-based neural networks have a strong capability of recognizing patterns on GUIs~\cite{Liu:2018:LDS:3242587.3242650, xi2019deepintent, chen2022towards}. In this paper, we train a ResNet-based network~\cite{he2016deep} (a variant of CNN-based network) to classify the detected graphical icons from the screenshots. Specifically, we employ ResNet-18 as the backbone and further enhance generalizability; i.e., we add dropout layers to every ResNet block to enhance its generalizability, preventing the potential over-fitting issue. We then train the ResNet model on the RICO-icon dataset with the default train-test split. After the training process, we apply the ResNet model to our dataset to obtain the corresponding data types of iconic GUI components as mentioned in Table~\ref{tab_keyword_list}.

\subsection{Segment Extraction Module}
Most mainstream app markets, such as the Google Play app store~\cite{googledeveloper} and the Apple App Store~\cite{appledeveloper}, require developers to provide a privacy policy link for their app on the homepage; hence, we can obtain those links based on a mobile app's name.
We employ \href{https://www.selenium.dev/}{\textit{Selenium}} and \href{https://www.crummy.com/software/BeautifulSoup/}{\textit{BeautifulSoup}}, two Python libraries, to capture these privacy policies in HTML format based on their privacy policy links.
To facilitate multilingual users and promote mobile apps in a global market, we observe that an increasing number of privacy policies are displayed in multiple languages.
We utilize a language detection library, namely \href{https://pypi.org/project/langdetect/}{\textit{langdetect}} to exclude non-English language texts from privacy policies to dispel ambiguity. Note that our framework can be easily extended to other languages.

Our ultimate goal is to obtain privacy policy segments with their corresponding privacy-related contexts.
According to the Flesch-Kincaid readability formula~\cite{kincaid1975derivation}, the readability of words is positively related to their length.
Document-level or paragraph-level privacy policies analysis is too coarse-grained for our task because we aim on retrieving the relevant sentence-based segments.
In addition, previous studies~\cite{torre2020ai, xie2022scrutinizing} have discussed that only using sentence-level privacy policies processing and analysis may lead to contradictory conclusions, since the same or similar sentences in different sections may have different implications.
Therefore, we adopt the multi-level privacy policies processing method proposed in~\cite{andow2019policylint, xie2022scrutinizing}.

For privacy policy documents whose HTML structure follow the $(\langle \texttt{Heading}\rangle \langle \texttt{Paragraph}\rangle ^ +)^+$ format, 
We reuse the pre-trained Bayesian multi-label classifier model~\cite{xie2022scrutinizing} to classify privacy policy paragraphs based on their headings since headings usually contain words that can be used as identifiers. 
The classifier is claimed to achieve 0.85 accuracy and can classify the majority of the documents based on their headings.
Paragraphs classified as \textit{Types} are related to the ``types of personal data collected by the current app''. Those paragraphs are highly related to the specific data practices, so we only conduct sentence-level analysis on them.


For privacy policy documents whose HTML structure does not follow the $(\langle \texttt{Heading}\rangle \langle \texttt{Paragraph}\rangle ^ +)^+$ format, we follow previous work~\cite{harkous2018polisis, windl2022automating} to classify paragraphs about specific data practices. Specifically, we train a CNN-based multi-label classification model on a large privacy policy dataset~\cite{wilson2016creation} with 23K fine-grained data practice annotations. Note that this dataset contains 12 high-level categories for data practices, including \textit{First-Party Collection/Use} and \textit{Third-Party Sharing/Collection}. We set the probability threshold at 0.5, i.e., privacy segments with predicted probability scores above 0.5 are classified as corresponding data practices. The results show we can achieve a 0.84 top-1 precision for classifying the privacy segments with their related data practices. 

After obtaining related privacy segments, we tokenize them into sentences by employing the \href{https://stanfordnlp.github.io/stanza/}{\textit{Stanza}} Python library.
We then conduct a two-stage sentence-level analysis to extract the sentences for constructing CPPs for each data type.
First, we perform a keyword search based on our collected keyword list (see Table~\ref{tab_keyword_list}).
For 1-gram keywords, we perform a case-insensitive word-wise comparison and check whether a keyword exists as sub-strings in a ``word'', e.g. \textit{location} and ``\textit{... your \underline{locationWe} may share...}'', since some punctuation or line break marks cannot be detected in rare cases. 
For keywords with two or more grams, we perform case-insensitive string-wise comparison, i.e., treating the target sentence as a string and searching identical n-gram keywords.
If a keyword exists, the sentence will be added to the corresponding privacy policy segments of its data type.
Second, for sentences without any keywords, we further employ a Bayesian binary classifier to determine whether the sentence is related to a data type. The classifier is trained on the dataset from \cite{xie2022scrutinizing}, reaching an accuracy of 0.98.
If yes, we use \href{https://github.com/explosion/spaCy}{\textit{SpaCy}} Python library to obtain noun chunks from the sentence. A noun chunk is a phrase that includes a noun and any connected words such as its adjectives.
We then calculate the phrases similarity $phrase\_sim( \, )$ between noun chunks and keywords, and the formula is:
\begin{equation*}
\label{formula_similarity}
\resizebox{.48\textwidth}{!}{%
    $phrase\_sim(p_1, p_2) = \frac{2 \times path\_similarity(p_1, p_2)}{word\_count({p_1}) + word\_count(p_2)}$
}%
\end{equation*}
where the $p_1$ and $p_2$ are a noun chunk and a keyword, respectively; $path\_similarity(\,)$ is the function from WordNet~\cite{miller1995wordnet}, reflecting the similarity based on parsing their semantic constituency trees' structure; $word\_count(\,)$ is to count the number of words.
If the similarity between a noun chunk and a keyword is higher than 0.8, an empirically set threshold, then we regard that the sentence that the noun chunk belongs to is the data type of the keyword.

Above all, we obtain the sentences related to each data type and group them as privacy policy segments. We also record the positions of keywords or noun chunks in the resulting sentences. For data types that do not match any sentences or paragraphs, we simply add "No relative information is found in the privacy policy." to notify users.

\subsection{CPPs Presentation Module}

For each privacy policy and its screenshot, we obtain their privacy policy segments and privacy contexts from the \textit{segments extraction module} and \textit{contexts detection module}, respectively. We then further process them for a more organized presentation to users.


First, we group the detected contexts of the same data types and map them to their corresponding privacy policy segments. For example, in Figure~\ref{fig_introduction}c, our framework, namely \textsc{SeePrivacy}, detects three contexts, including two textual GUI components, i.e., \textit{``Share your location''} and \textit{``use your birthday,''} and one iconic GUI components, such as a location mark icon in the right middle of the screenshot. SeePrivacy classifies them as \textit{Location}, \textit{Location}, and \textit{Birthday} data types individually. 
The text \textit{``Share your location''} and the location mark icon are coupled with the segment about location information. The text \textit{``use your birthday''} solely corresponds to the segment about \textit{Birthday} data type.
A previous study~\cite{palmen2023bold} points out that appropriately setting bold fonts for text will significantly increase its readability. Hence, for the privacy policy segments displayed to the users, we render the keywords or noun chunks in bold fonts based on their positions.

\section{Evaluation and Discussion}
\label{sec_evaluation}

\begin{table*}
\caption{Performance of SeePrivacy's modules. CDM stands for Context Detection Module.}
\vspace{-5pt}
\label{tab_evaluation_total}
\subcaptionbox{CDM - Textual GUI components}{
\resizebox{0.2655\textwidth}{!}{%
    \begin{tabular}{lccc}
        \toprule
        \textbf{Category} & \textbf{Accuracy} & \textbf{Precision} & \textbf{Recall} \\
        \midrule
        Name & 0.92 & 0.92 & 1.00 \\
        Birthday  & 1.00 & 1.00 & 1.00 \\
        Address  & 0.38 & 0.43 & 0.79 \\
        Phone  & 0.90 & 0.96 & 0.93 \\
        Email  & 0.67 & 0.97 & 0.68 \\
        Profile  & 0.45 & 0.47 & 0.94 \\
        Contacts  & 0.95 & 0.98 & 0.98 \\
        Location  & 0.89 & 0.95 & 0.94 \\
        Photos  & 0.78 & 0.97 & 0.80 \\
        Voices  & 1.00 & 1.00 & 1.00 \\
        Financial info  & 0.82 & 0.93 & 0.88 \\
        Social media  & 0.60 & 0.75 & 0.75 \\
        \midrule
        \textbf{Average} & 0.78 & 0.86 & 0.89 \\
        \bottomrule
    \end{tabular}
}%
}
\hfill
\subcaptionbox{CDM - Iconic GUI components}{
\resizebox{0.328\textwidth}{!}{%
\begin{tabular}{llrrr}
\toprule
\textbf{RICO-icon Class} & \textbf{Category} & \textbf{Accuracy} & \textbf{Precision} & \textbf{Recall} \\
\midrule
Call  & Phone & 0.70 & 1.00 & 0.70 \\
Email  & Email & 0.74 & 0.68 & 0.81 \\
Avatar  & Profile & 0.85 & 0.82 & 0.89 \\
Follow & Contacts & 0.90 & 0.82 & 1.00 \\
Group  & Contacts &  0.67 & 0.73 & 0.62   \\
Location  & Location & 0.89 & 0.86 & 0.92 \\
Location crosshair & Location & 0.72 & 0.63 & 0.83 \\
Photo & Photos & 0.87 & 0.89 & 0.85 \\
Wallpaper  & Photos & 0.87 & 0.82  & 0.93 \\
Videocam  & Photos &  0.91 & 0.83 & 1.00 \\
Microphone  & Voices & 0.96 & 0.93  & 1.00   \\
Cart &  Financial Info & 0.81 & 0.75   &   0.88   \\
Facebook  & Social media & 0.83 & 0.75 & 0.92  \\
Twitter  & Social media & 0.90  & 0.95 & 0.86 \\
\midrule
\textbf{Average} & &  0.82 & 0.83 & 0.85  \\
\bottomrule
\end{tabular}
}%
}
\hfill
\subcaptionbox{Segments Extraction Module}{
    \resizebox{0.3465\textwidth}{!}{%
    \begin{tabular}{lcccc}
    \toprule
    \textbf{Category}    & \textbf{Accuracy} & \textbf{Precision} & \textbf{Recall} & \textbf{Success Rate}\\
    \midrule
    Name        & 0.98 & 0.97 & 1.00 & 0.92 \\
    Birthday    & 0.94 & 1.00 & 0.94 & 0.62 \\
    Address     & 0.48 & 0.82 & 0.50 & 0.78 \\
    Phone       & 1.00 & 1.00 & 1.00 & 0.97 \\
    Email       & 1.00 & 1.00 & 1.00 & 0.98 \\
    Profile     & 0.96 & 0.96 & 1.00 & 0.97 \\
    Contacts    & 1.00 & 1.00 & 1.00 & 1.00 \\
    Location    & 0.92 & 0.94 & 0.98 & 0.91 \\
    Photos      & 0.84 & 0.83 & 1.00 & 1.00 \\
    Voices      & 0.85 & 0.33 & 1.00 & 1.00 \\
    Financial info   & 1.00 & 1.00 & 1.00 & 0.96 \\
    Social media& 1.00 & 1.00 & 1.00 & 1.00 \\
    \midrule
    \textbf{Average} & 0.92 & 0.94 & 0.98 & 0.96  \\
    \bottomrule
    \end{tabular}
    }%
}
\vspace{-5pt}
\end{table*}

In this section, we mainly investigate and discuss the following two research questions:
\begin{itemize}[leftmargin=*]
\item \textbf{RQ1:} What is the performance of SeePrivacy in automatically generating contextual privacy policies for mobile apps?
\item \textbf{RQ2:} What is the usability of SeePrivacy in practice?
\end{itemize}

\subsection{RQ1: Performance Evaluation}
As our framework is composed of three main modules, we evaluate the performance of each of them on our benchmark dataset, \textsc{Cpp4App}. 
\subsubsection{Context Detection Module}
To determine whether a detected GUI component matches the ground truth, we introduce Intersection over Union (IoU), which is a commonly used evaluation metric to measure the performance of an object detection model.
If the IoU between ground truth and the detected component area is greater than a threshold $\beta$, then we regard it as a match. We set $\beta = 0.5$.


\textbf{Textual GUI components.} Table~\ref{tab_evaluation_total}a presents the findings regarding the detection of textual GUI components. In general, our approach demonstrates a commendable ability to accurately recognize and classify texts on mobile GUI screenshots, achieving an accuracy of 0.78, precision of 0.86, and recall of 0.89.
Regarding specific categories, certain categories, such as \textit{Address}, exhibit performance below the average level due to their susceptibility to being misclassified as semantically similar categories, like \textit{Location}. 
Furthermore, it is noteworthy that the accuracy metric yields considerably lower results compared to the other metrics.
To investigate the underlying causes, we select a random sample comprising 5\% of screenshots from the benchmark dataset and manually assess the performance of PP-OCR. The results indicate an H-mean score of over 0.98 for text detection and an accuracy of over 0.98 for text recognition.
Additionally, we conduct a detailed analysis of the failure cases of chatGPT, which reveals its inclination to classify a piece of text as "not relevant". We observe that when we exclude the descriptions of each data type from the chatGPT, the occurrence of "not relevant" diminishes, but this change also leads to a significant decline in overall performance.

\begin{figure}[!t]%
    \centering
    \subfloat[\centering Screenshot 14-2 ]{{\includegraphics[width=0.15\textwidth]{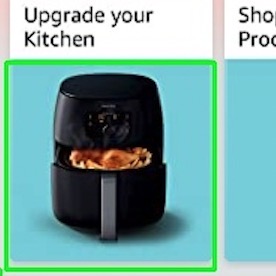}}}%
    \qquad
    \subfloat[\centering Screenshot 20-3]{{\includegraphics[width=0.15\textwidth]{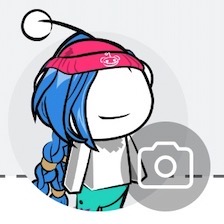}}}%
    \caption{Failure cases of the mislabel (left) and the missing label (right).}%
    \label{fig:example}%
\vspace{-5pt}
\end{figure}

\textbf{Iconic GUI components.}
Table~\ref{tab_icon_deteciton_comparison} shows the comparison of the icons localization capability between our method and previous methods.
In our experiment, we manually count the number of icons from the benchmark dataset and compare them. 
Results show that our proposed method has better performance in identifying icons curated on GUI screenshots. We manually examine failure cases and find that most of the missing labels are caused by overlapping with other components or interleaved boundaries. As for the mislabelled cases, small squarish images, e.g., item images in shopping apps, tend to be wrongly included as an icon. Fig.~\ref{fig:example} shows two examples of failure cases in our experiment. 

\begin{table}[h]
\centering
\caption{Performance comparison between iconic GUI components localization methods.}
\label{tab_icon_deteciton_comparison}

\begin{tabular}{lcc}
\toprule
\textbf{Method}   & \textbf{Precision (mislabel)} & \textbf{Recall (missing label)} \\
\midrule
UIED~\cite{xie2020uied}  &0.78 & 0.44\\
Liu et al.~\cite{Liu:2018:LDS:3242587.3242650}   & 0.79 &0.86 \\
Ours  &\textbf{0.84 }&\textbf{0.88}  \\
\bottomrule
\end{tabular}
\vspace{-5pt}
\end{table}


We also compare the icon classification models on the RICO-icon dataset. Table~\ref{tab_classification_comparison} shows that our model achieves competitive performance but much better ($\sim$14$\times$) efficiency compares to the baseline.
Moreover, the class-wise breakdown is shown in Table~\ref{tab_evaluation_total}b.

\begin{table}[h]
\centering
\caption{Performance and efficiency comparison for iconic GUI elements classification on the RICO-icon dataset. Time(s) represents the elapsed seconds to process a thousand icons. }
\label{tab_classification_comparison}
\vspace{-5pt}
\begin{tabular}{lcccc}
\toprule
\textbf{Model}                        & \textbf{Accuracy} & \textbf{Precision} & \textbf{Recall} & \textbf{Time(s)} \\
\midrule
Liu et al.~\cite{Liu:2018:LDS:3242587.3242650}   & 0.90       & \textbf{0.90}       & 0.79   & 5.02  \\
Ours                    & \textbf{0.91}       &0.80        & \textbf{0.81}   & \textbf{0.37}\\
\bottomrule
\end{tabular}
\vspace{-5pt}
\end{table}


\begin{table*}[h]
\centering
\caption{Questions for the user study. Except $\textit{Q}_2$ and $\textit{Q}_9$, all other questions are on a 5-point Likert scale.}
\vspace{-5pt}
\label{tab_questions}
\resizebox{.95\textwidth}{!}{%
\begin{tabular}{l|l|l}
\toprule
No. & Question & Scale\\

\midrule
$\textit{Q}_1$  &How concerned are you about your privacy information while using mobile apps? & 5 for very concerned, 1 for very unconcerned\\
$\textit{Q}_2$   & Have you been troubled by the privacy related issues when using mobile apps? & Yes, No, or Maybe \\
$\textit{Q}_3$    & Do you read mobile app's privacy policies when you encountered? &  5 for always read, 1 for never read \\

\midrule

$\textit{Q}_4$  & Does the data type match with the detected context? & 5 for strongly agree, 1 for strongly disagree\\
$\textit{Q}_5$   & Does the data type match with the displayed privacy policy segment? & 5 for strongly agree, 1 for strongly disagree \\
$\textit{Q}_6$    &  Does the displayed privacy policy segment explain the detected context on the screenshot? & 5 for strongly agree, 1 for strongly disagree \\

\midrule
$\textit{Q}_7$   & What do you think the usefulness of this tool in terms of providing privacy information for mobile apps? & 5 for very useful, 1 for very useless\\
$\textit{Q}_8$  & Will you read contextual privacy policies when you encountered in future when you encountered? & 5 for always read, 1 for never read \\
$\textit{Q}_9$   & What suggestions do you have for this tool? & Completion question\\
\bottomrule
\end{tabular}
}%
\end{table*}


\begin{figure*}[!h]%
    \centering 
    \includegraphics[width=.95\textwidth]{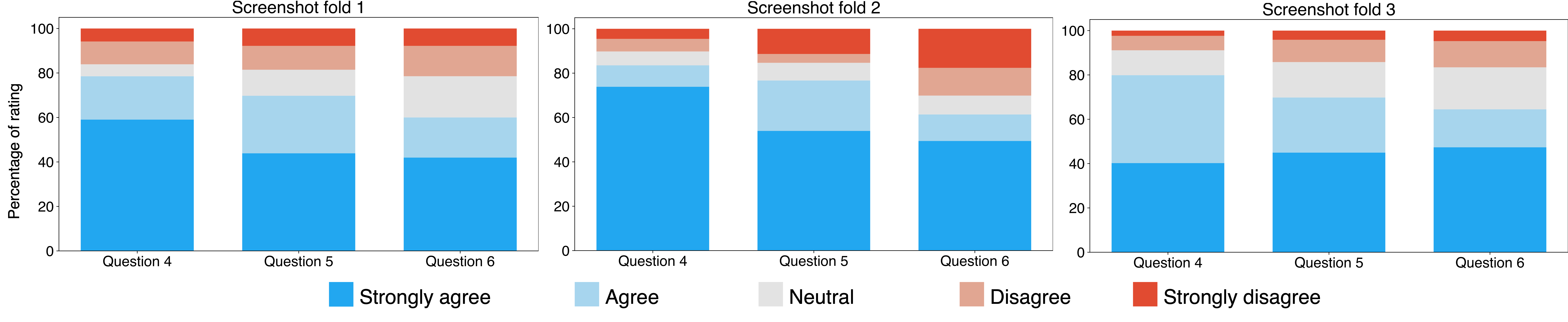}
    \vspace{-5pt}
    \caption{The percentage values of participants’ agreement to our three statements on a 5-point Likert scale.}
    \label{fig_user_study}
\end{figure*}

\textbf{Overall context identification.}
%
We also combine the textual and iconic GUI components and evaluate the overall context to better understand the performance of our proposed framework. Table~\ref{tab:performance_metrics} shows the overall context identification results considering textual and iconic GUI components. Specifically, SeePrivacy can achieve 0.76, 0.86, and 0.87 in terms of accuracy, precision, and recall, respectively. We also estimate the performance of overall context identification as per a screenshot, denoted by a context coverage rate. Specifically, the context coverage rate calculates how many data types the contexts belong to that can be detected by our method. For example, in Fig.~\ref{fig_context_cover_rate}, there is one iconic GUI component (green bounding box) and three textual GUI components (red bounding boxes). Since all those components belong to \textit{Location}, if any of those components are not detected, the context coverage rate is still 1.0 for this screenshot. Even if our framework fails to detect some contexts, users can still read their corresponding privacy policy segments via other contexts. Overall, our method yields an \textbf{83.6\%} context coverage rate. 

\begin{table}
	\begin{minipage}{0.55\linewidth}
		\caption{\small Category-wise overall context identification.}
		\label{tab:performance_metrics}
		\vspace{-5pt}
		\centering
\resizebox{1\linewidth}{!}{%
\begin{tabular}{lccc}
\toprule
Category  & Accuracy  & Precision & Recall  \\
\midrule
Name        & 0.92        & 0.92     & 1.00       \\
Birthday    & 1.00        & 1.00     & 1.00       \\
Address     & 0.38        & 0.43     & 0.79       \\
Phone       & 0.85        & 0.91     & 0.94       \\
Email       & 0.67        & 0.90     & 0.72       \\
Profile     & 0.62        & 0.64     & 0.95\\
Contacts    & 0.87        & 0.98     & 0.88       \\
Location    & 0.84        & 0.95     & 0.87       \\
Photos      & 0.78        & 0.95     & 0.81       \\
Voices  & 0.89       & 1.00     & 0.89       \\
Financial info   & 0.74        & 0.93     & 0.78       \\
Social media& 0.60        & 0.75     & 0.86       \\
\midrule
\textbf{Average} & 0.76 & 0.86 & 0.87 \\
\bottomrule
\end{tabular}
}%
\end{minipage}
	\hfill
	\raisebox{-0.035\totalheight}{%
	\begin{minipage}{0.4\linewidth}
		\centering
		\includegraphics[width=0.97\linewidth]{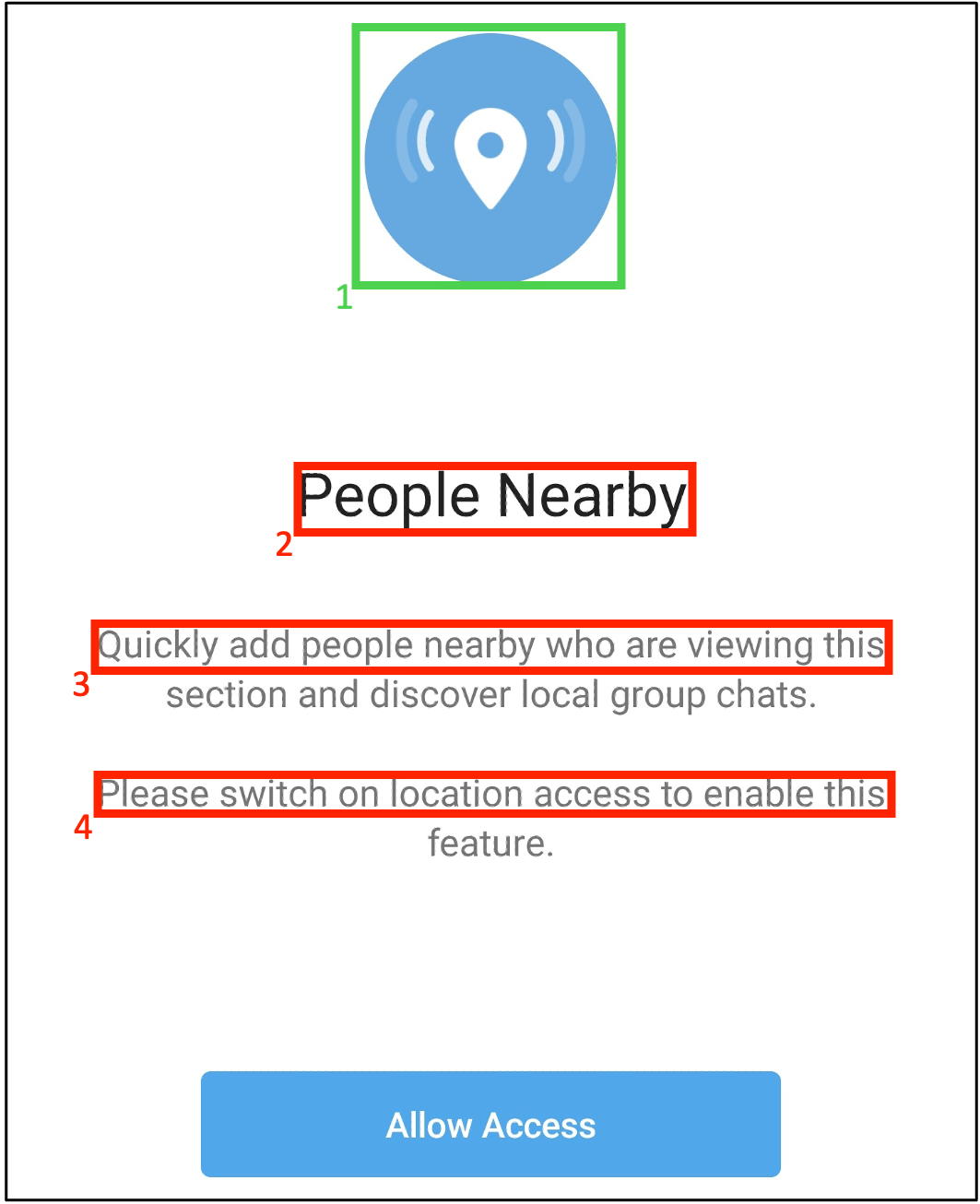}
		\vspace{-5pt}
		\captionof{figure}{Screenshot 17-2}
		\label{fig_context_cover_rate}
	\end{minipage}
	}%
\vspace{-5pt}
\end{table}

\subsubsection{Segment Extraction Module}
Table~\ref{tab_evaluation_total}c shows the results of the segment extraction module.
In the benchmark dataset, some mobile apps' data types are not relevant to any privacy policy segment (``No relative information is found in the privacy policy''), so the framework is expected to not retrieve anything either. Thus, the classification metrics (accuracy, precision, and recall) reflect privacy policy segment retrieval performance for each data type. 
The success rate denotes how many retrieved privacy policy segments match their ground truth. Specifically, the majority of a successfully retrieved privacy policy segment is identical to the ground truth, thus we introduce the segment similarity $segment\_sim(\,)$ to decide whether retrieval is successful.
For given ground truth segment $s_{gt}$ and retrieved segment $s_{ret}$:

\vspace{-12pt}
\label{formula_similarity}
\begin{align*}
    segment\_sim(s_{ret}, s_{gt}) &= \frac{1}{\textsc{Min}(n, m)}\sum_{i=1}^n\sum_{j=1}^m \overline {lcs}(p^{ret}_{i}, p^{gt}_{j}) \\ 
    \overline {lcs}(p^{ret}_{i}, p^{gt}_{j})  &=\frac{lcs(p^{ret}_{i}, p^{gt}_{j})} {\textsc{Min}(\textsc{Len}(p^{ret}_{i}), \textsc{len}(p^{gt}_{j}))}
\end{align*}
\vspace{-2pt}

where $lcs(\,)$ stands for the function to calculate longest common string; $p^{ret}_{j}$ are phrases separated by punctuation of retrieved segments and $p^{gt}_{i}$ are labelled phrases of ground truth segments. If it is greater than 0.8, then the segment pair is regarded as a successful match.

From the results, we can observe that our framework generally demonstrates strong performance in retrieving not only the privacy policy segments but the relevant ones. There are some categories, such as \textit{Address}, which are more complex than the other categories due to relatively ambiguous keywords. In addition, there are too little data (less than 5 records) for \textit{Voices} according to the benchmark dataset, thus the limited performance of this category could be caused by the randomness.

\subsection{RQ2: User study}

To validate the functionality and demonstrate the usefulness of our tool in practice, we conduct a usability evaluation to comprehensively examine SeePrivacy for generating contextual privacy policies for mobile apps.

\subsubsection{User study design} 
Table~\ref{tab_questions} lists all the questions used in our user study. We recruit participants based on the following criteria: being over 18 years old, fluent in English reading, and capable of using mobile apps without special accessibility tools (e.g., VoiceOver on iOS) via mailing lists of authors' institutions. In addition to the general introduction and consent information of this project, we also provide descriptions of the data types, specific instructions, and an example.
Our questions cover three aspects: privacy policies reading habits for mobile apps ($\textit{Q}_{1\text{-}3}$), specific usability evaluation ($\textit{Q}_{4\text{-}6}$), and general usability evaluation ($\textit{Q}_{7\text{-}9}$) of the SeePrivacy. 

As for the specific usability evaluation, we randomly sample 30 screenshots with their privacy policies from the benchmark dataset and employ SeePrivacy to generate CPPs.
We then divide the generated CPPs into three folds and compose the questionnaire for each fold, i.e., ten screenshots in each questionnaire as manual examination involves very heavy privacy policy reading and comprehending tasks.
Then the participants need to answer $\textit{Q}_{4\text{-}6}$ for all the ten allocated screenshots, and they are asked to rate the agreement level on a 5-point Likert scale ranging from five (strongly agree) to one (strongly disagree).

\subsubsection{User study results}

In total, we recruit 15 people for the study (6 females, 8 males, and 1 who preferred not to be mentioned). Each screenshot fold is evaluated by five people. We first analyze how participants rate the statements in a specific usability evaluation. Figure~\ref{fig_user_study} and Table~\ref{tab_user_study_results} show the results of our user study. The matching between data types and detected contexts received high ratings, with over 81\% of their responses being either ``Agree'' or even ``Strongly Agree,'' and its values for the mean (4.22) and median (5) are also high. As for data types and privacy policy segments, over 72\% of participants gave positive ratings, and their mean and median are 3.95 and 4, respectively. More than 77\% of participants do not disagree that retrieved privacy policy segments explain detected contexts, which is underlined by a 3.75 mean and a 4 median. Overall, SeePrivacy exhibits excellent functionality to deliver contextual privacy policies to users.


According to our user study, we found that people are concerned about their privacy information while using mobile apps (4.13 mean and 4 median for $\textit{Q}_1$). However, they are commonly disturbed by issues about privacy information (13 out of 15 answered ``Yes'' or ``Maybe'' for $\textit{Q}_2$). The usefulness of SeePrivacy is highly regarded in terms of providing privacy information on mobile apps. 14 out of 15 participants rated ``Very useful'' or ``Useful''. Moreover, we also compare the reading habits of common privacy policies and CPPs. Results show that people are much more willing to read CPPs (a mean increase from 2.00 to 4.07), indicating that our SeePrivacy is essential and significant in helping people overcome the ``privacy policy reading phobia''. 

\begin{table}[!t]
\centering
\caption{Statistic results of our user study. SD stands for standard deviations.}
\label{tab_user_study_results}
\vspace{-5pt}
\resizebox{.8\linewidth}{!}{%
\begin{tabular}{llrcc}
\toprule
\textbf{No.} & \textbf{Topic} & \textbf{Mean}  & \textbf{Median}  & \textbf{SD}\\
\midrule
$\textit{Q}_4$  & Data type \& Context&  4.22 & 5 & 1.14\\
$\textit{Q}_5$  & Data type \& Policy segment& 3.95 & 4& 1.28\\
$\textit{Q}_6$  & Policy segment \& Context& 3.75 & 4 & 1.40\\
\midrule
\midrule
$\textit{Q}_1$  & Privacy concerns (-)    & 4.13 & 4 &  1.09\\
$\textit{Q}_7$  & CPP usefulness (+)  & 4.53 & 5 & 0.62 \\
\midrule
\midrule
$\textit{Q}_3$  & PP Reading frequency    &2.00 & 2 & 1.03 \\
$\textit{Q}_8$  & CPP Reading frequency    & 4.07 & 4 & 0.85 \\
\midrule
\multicolumn{2}{l}{\textbf{Improvement}} & \textbf{+2.07} & \textbf{+2} & / \\
\bottomrule
\end{tabular}
}%
\vspace{-15pt}
\end{table}

\section{Threats to Validity}
\label{sec_threats}

\textbf{Internal validity.} Threats to internal validity refer to errors in our experiments.
We reuse several pre-trained classifiers to process privacy policies in the paragraph- and sentence-level analyses. We employ PP-OCR and chatGPT for textual GUI element detection. We also follow OpenAI’s API to access chatGPT. We specifically evaluate the performance of every module on the benchmark dataset and conduct the error analysis. 
We also propose a user study to further validate the functionality and usability of our framework in practice. We have rechecked the code and data, but errors may still exist.

\textbf{External validity.} Threats to external validity refer to the generalizability of our proposed method. In our experiments, although our benchmark dataset covers various application categories, it is mainly composed of relatively popular mobile applications in each category. However, we maneuver a set of selection rules to collect different pages in applications as completely as possible. 
In addition, our method is based on learning-based GUI pattern understanding and privacy policy analysis, which is well-known for its generalizability. Thus, we believe that there is a minimal threat to external validity.
\section{Related Work}
\label{sec_relatedwork}

\textbf{Vision-based mobile GUI understanding.}
Previous researchers have been intensively studying this topic from different perspectives.
White et al.~\cite{white2019improving} leverage a famous object detection model called YOLOv2, to detect GUI widgets in-app screenshots for testing. 
Chen et al.~\cite{chen2019gallery} apply Faster R-CNN to obtain GUI widgets from app screenshots to construct a GUI widget gallery. 
In another work, Chen et al.~\cite{chen2020object} propose a hybrid method that combines traditional rule-based and learning-based CV techniques, achieving the start-of-the-art performance for GUI widget detection.
Xie et al.~\cite{xie2022psychologically} present a novel unsupervised image-based method for inferring perceptual groups of GUI widgets. 

\textbf{Privacy policy analysis.}
Wilson et al.~\cite{wilson2016creation} build a privacy policy corpus with 23K fine-grained data practice annotations.
Kaur et al.~\cite{kaur2018comprehensive} perform a comprehensive analysis of keywords and content of over two thousand online policy documents.
Zimmeck et al.~\cite{zimmeck2019maps} introduce the
Mobile App Privacy System (MAPS) for conducting an
extensive privacy census of Android apps. The evaluation of a million Android apps reveals policies that do exist are often silent on the practices performed by apps.
Amos et al.~\cite{amos2021privacy} curate a comprehensive dataset of a million privacy policies spanning over two decades, and their analyses ``paint a troubling picture'' of the privacy policy's transparency and accessibility.

The most closet work is~\cite{windl2022automating}, called PrivacyInjector, used to automatically generate contextual privacy policies for websites. However, their context detection method is intrusive and based on ad-hoc HTML analysis, which is not applicable to mobile applications. In addition, their definitions of privacy-related contexts are more coarse-grained compared to this work and specific to website scenarios.
\section{Conclusion and Future Work}

Privacy policies are extensively criticized due to their poor readability. Contextual privacy policies are proposed to solve this research challenge. We first construct the benchmark dataset, namely \textsc{Cpp4App}, containing over 600 privacy-related contexts with their corresponding privacy policy segments in the mobile scenario, based on our comprehensive survey. We then present a novel framework, called \textsc{SeePrivacy}, to automatically generate contextual privacy policies for mobile apps to respond to this need. We utilize computer vision techniques and a pre-trained Large Language Model, i.e., chatGPT, to detect privacy-related contexts on mobile GUI screenshots and leverage NLP techniques to extract corresponding segments from the privacy policy document. Our evaluation results show that our method yields an 83.6\% coverage rate for privacy-related context detection and an accuracy of 0.92 for extracting corresponding policy segments. Moreover, 96\% of the retrieved policy segments can be correctly matched with their contexts. Additionally, we perform a user study to further validate the functionality and usability in practice. The results show that more than 77\% of participants do not disagree that retrieved privacy policy segments explain detected contexts, which is underlined by a 3.75 mean and 4 median on a 5-point Likert scale. The usefulness rating received an average of 4.53 out of 5 from participants. We believe that our work provides a new research direction for the community, sheds light on developing contextual privacy policies for mobile apps, and effectively assists users in comprehending privacy notices.

The promising results of this work illuminate two potential directions for future research. First, some participants ($\textit{P}_{7}$ and $\textit{P}_{15}$) suggested that aside from CPPs, they also care about general data rights such as the Right To Be Forgotten or Right to Control. We plan to extend our framework to be capable of extracting and presenting data rights information from privacy policies. 
Second, although our original intention is to deliver the original privacy policy content without any modification, some retrieved policy segments are too long to comprehend in a timely manner. We believe it is worthwhile to evaluate how to digest (e.g., by pre-trained Large Language Models) lengthy policy segments with minimum information loss and distortion to maintain their integrity and practicability simultaneously.

\textbf{Data Availability.}
Our dataset and codes are available at: \url{https://github.com/Cpp4App/Cpp4App}.


\bibliographystyle{IEEEtran}
\bibliography{9_References}

\end{document}